\documentclass[
  prd,
  aps,
  superscriptaddress,
  showpacs,
  twocolumn,
  nofootinbib
]{revtex4-2}

\usepackage[T1]{fontenc}
\usepackage{newtxtext}
\usepackage{newtxmath}
\usepackage{mathrsfs}

\usepackage{amsmath}
\usepackage{mathtools}
\usepackage{slashed}
\usepackage{braket}

\usepackage{graphicx}
\usepackage[dvipsnames]{xcolor}

\definecolor{myred}{rgb}{0.9,0,0}
\definecolor{navy}{rgb}{0.05,0.05,0.8}

\usepackage[colorlinks=true,citecolor=myred,linkcolor=navy,urlcolor=navy]{hyperref}
\usepackage{cleveref}

\usepackage{enumitem}
\usepackage{setspace}

\usepackage{xspace}
\usepackage{xstring}

\newcommand{\be}{\begin{equation}}
\newcommand{\ee}{\end{equation}}
\newcommand{\bs}{\begin{split}}
\newcommand{\es}{\end{split}}

\newcommand{\eq}[1]{Eq.~\eqref{#1}}

\begin{document}

\title{Linear monitoring-feedback Model for Linearized gravity}

\author{Daisuke Miki}
\affiliation{Burke Institute for Theoretical Physics, California Institute of Technology, Pasadena, California 91125}
\author{Alfred Li}
\affiliation{Burke Institute for Theoretical Physics, California Institute of Technology, Pasadena, California 91125}
\author{Yiqiu Ma}
\affiliation{National Gravitational Laboratory, Hubei Key Laboratory of Gravitation and Quantum Physics, School of Physics, Huazhong University of Science and Technology, Wuhan, 430074, China}
\author{Yubao Liu}
\affiliation{State Key Laboratory of Low Dimensional Quantum Physics,
Department of Physics, Tsinghua University, Beijing, 100084, China}
\author{Yanbei Chen}
\affiliation{Burke Institute for Theoretical Physics, California Institute of Technology, Pasadena, California 91125}

\begin{abstract}
Whether gravity has quantum degrees of freedom is a question of fundamental importance. We formulate a linear monitoring-feedback framework for gravitational interactions that includes local classical-record models of semiclassical gravity and, when quantum resources are supplied in correlated ancilla inputs, can reproduce linearized quantum-gravity dynamics. Auxiliary systems monitor the positions of massive bodies, and their records are fed back through a chosen gravitational response kernel. At the level of linear input-output theory, this construction makes the deterministic response explicit while exposing the force noise and correlations associated with the auxiliary sector.
We organize the allowed ancilla correlations into three regimes. In Regime~I, independent ancillas have unsqueezed, uncorrelated sensing and back-action quadratures. A single object then has a minimum classical-gravity noise (CGN), $S_{\mathscr{F}}^{\rm min}=2\hbar M\omega_{\rm SN}^2$, where \(\omega_{\rm SN}\) is the Schr\"odinger--Newton frequency set by the microscopic mass distribution.
In Regime~II, arbitrary local sensing--back-action correlations are allowed, but the ancillas associated with different objects remain independent. The single-object CGN is then unrestricted, while the joint force-noise matrix obeys a determinant bound that is equivalent to the local classical-channel/no-entanglement bound. In Regime~III, cross-object correlations and a common nonlocal mediator are allowed. The commuting force noises can then be jointly squeezed, entanglement between the masses can occur, and—with a suitable retarded response and input state—the monitoring-feedback system can reproduce the noise and commutators of linearized quantum gravity, including non-Markovian quantum fluctuations.
The framework therefore provides a common language for comparing distinct linear gravity theories and for designing experiments that test gravity's quantum nature. Entanglement witnesses, force-noise spectra, cross-correlations, and finite-propagation effects become complementary diagnostics for determining which classes of classical or quantum mediators remain compatible with an observed interaction.
\end{abstract}

\maketitle

\section{Introduction}

Testing whether gravity behaves quantum mechanically is one of the central experimental problems at the interface of quantum mechanics and gravity.  A variety of proposals have been put forward to search for gravitational effects in quantum systems, including gravity-induced entanglement~\cite{bose2017spin,marletto2017gravitationally}, gravity-induced decoherence~\cite{diosi1987universal,penrose1996gravity,bassi2013models,bassi2023collapse}, self-gravity effects, and signatures associated with propagation or fluctuations of the gravitational field~\cite{belenchia2018quantum,christodoulou2023locally,mitrakos2026does}.  These proposals are not based on a single theoretical hypothesis: some assume that gravity is classical, and ask how it can act consistently on quantum matter and predicts the possible existence of additional gravity-induced-noise;  whereas others assume a quantized gravitational field and ask which of its quantum features can be observed.  A useful framework should make the difference of these models explicit at the level of experimentally accessible observables.

Kafri, Taylor, and Milburn (KTM) provided a particularly useful starting point by realizing a classical Newtonian interaction as a continuous monitoring-feedback process~\cite{kafri2014classical}.  In the KTM model, each object is coupled to an auxiliary observer; the observer produces a classical record, and a chosen response kernel feeds that record back onto the masses and generates gravitational interaction.  Our construction in this work makes clear how the local and independent ancillas give semiclassical classical-channel models, whereas nonlocal quantum correlations in the inputs of the ancillas extend the same linear framework to quantum-mediated interactions.

The main results are organized by the correlations available in the auxiliary sector.  In Regime\,I, independent uncorrelated monitoring gives a single-body self-gravity noise floor $S_{\mathscr F}^{\rm min}=2\hbar M\omega_{\rm SN}^2$, with $\omega_{\rm SN}$ set by microscopic mass localization; this is a natural first target for experiments. For crystalline solids, treating the internal matter distribution as part of the classical record makes this scale much larger than the mutual-gravity scale relevant to two-body entanglement proposals. In Regime\,II, local correlations can redistribute the noise, but the joint force-noise matrix obeys the local classical-channel determinant bound and cannot generate gravity-induced entanglement.  In Regime\,III, nonlocal quantum correlations between ancilla inputs can jointly squeeze the commuting force noises and generate entanglement.  With a suitable causal response kernel, their commutators and fluctuations can moreover be chosen to coincide with those of a linearized quantum-gravity field. Force-noise measurements, cross-correlations, entanglement witnesses, and propagation-delay effects therefore provide complementary tests of the assumptions behind a gravitational interaction, which our generalized linear monitoring-feedback model can cover.

The structure of this paper is organized as follows. Section~\ref{sec:modeloverview} introduces the monitoring--feedback architecture and the hierarchy of auxiliary correlations. Section~\ref{sec:single} then analyzes self-gravity for one object, establishing the single-body noise scale and its dependence on the local auxiliary input. Section~\ref{sec:twoobjects} extends the analysis to two mutually gravitating objects while retaining independent, local ancillas, and derives the local determinant bound underlying the no-entanglement result. Section~\ref{sec:experiments} translates these two local-record regimes into experimental targets. Finally, Sec.~\ref{sec:RIII} allows nonlocal correlations between the auxiliary inputs: it first treats the instantaneous Newtonian limit, and then extends the construction to a retarded response and its operational connection to linearized quantum gravity. Section~\ref{sec:conclusions} summarizes the implications for tests of gravity's quantum nature.

\section{Model framework}
\label{sec:modeloverview}

In a semiclassical gravity theory, the gravitational field is sourced by {\it classical information} about the mass density of quantum matter.  The relevant questions are therefore what counts as a classical record, how that record is obtained, and what fluctuations are introduced in obtaining it.  Here, by classical information we mean the eigenvalues of a set of commuting observables.  Since these observables are simultaneously measurable~\cite{braginsky1995quantum}, their eigenvalues can be represented as the record of a measurement process.  At the level of an enlarged system, the record-extraction step can use only unitary evolution and Born's rule and can therefore be embedded in a larger quantum model~\cite{tomizuka2026emergence}; this dilation does not, by itself, turn the gravitational channel into a quantum mediator.

In earlier work~\cite{miki2025role,liu2023semiclassical,liu2025semiclassical,liu2026testing}, the classical information driving gravity was obtained by the experimentalist within the Causal-Conditional Schr\"odinger--Newton (CCSN) theory.  CCSN restores causality and energy-momentum conservation to Schr\"odinger--Newton theory while retaining experimentally meaningful signatures~\cite{Restoring2026}.  Because no additional ancilla sector is used to extract the source information, CCSN introduces no monitoring fluctuations beyond those already required by the experiment. In contrast to CCSN, this work focuses on sourcing classical gravity from information obtained by additional, auxiliary observers/ancillas, not by the experimentalists. The noise associated with these auxiliary observers becomes one phenomenological effect that can be experimentally searched.

This auxiliary-observer architecture underlies the classical-channel model of KTM~\cite{kafri2014classical}, the broader Classical-Quantum framework of Oppenheim and collaborators~\cite{oppenheim2023postquantum,oppenheim2023gravitationally}, and the Newtonian master equation of Kryhin and Sudhir~\cite{kryhin2025distinguishable}.  Fabiano \emph{et al.}~\cite{fabiano2026minimal} derived the corresponding model-independent noise and entanglement thresholds.  The same family is connected to the Di\'osi--Penrose model and other collapse models~\cite{diosi1987universal,penrose1996gravity,bahrami2014schrodinger,nimmrichter2015stochastic,tilloy2016sourcing}.  Tomizuka and Takeda~\cite{tomizuka2026emergence} and Hotta \emph{et al.}~\cite{hotta2026} further showed how completely positive CQ models can be embedded in a larger quantum theory. Therefore, a general model constructed in this work should be broad enough to cover the above different models, and its ingredients should include observers, their input states and interactions with the masses, the classical information produced by the measurement process, and the feedback of this classical information that generates the gravitational response. Furthermore, a sufficiently general model should even be able to describe linear quantum gravity under certain conditions.

\begin{figure*}[t]
    \includegraphics[width=0.8\textwidth]{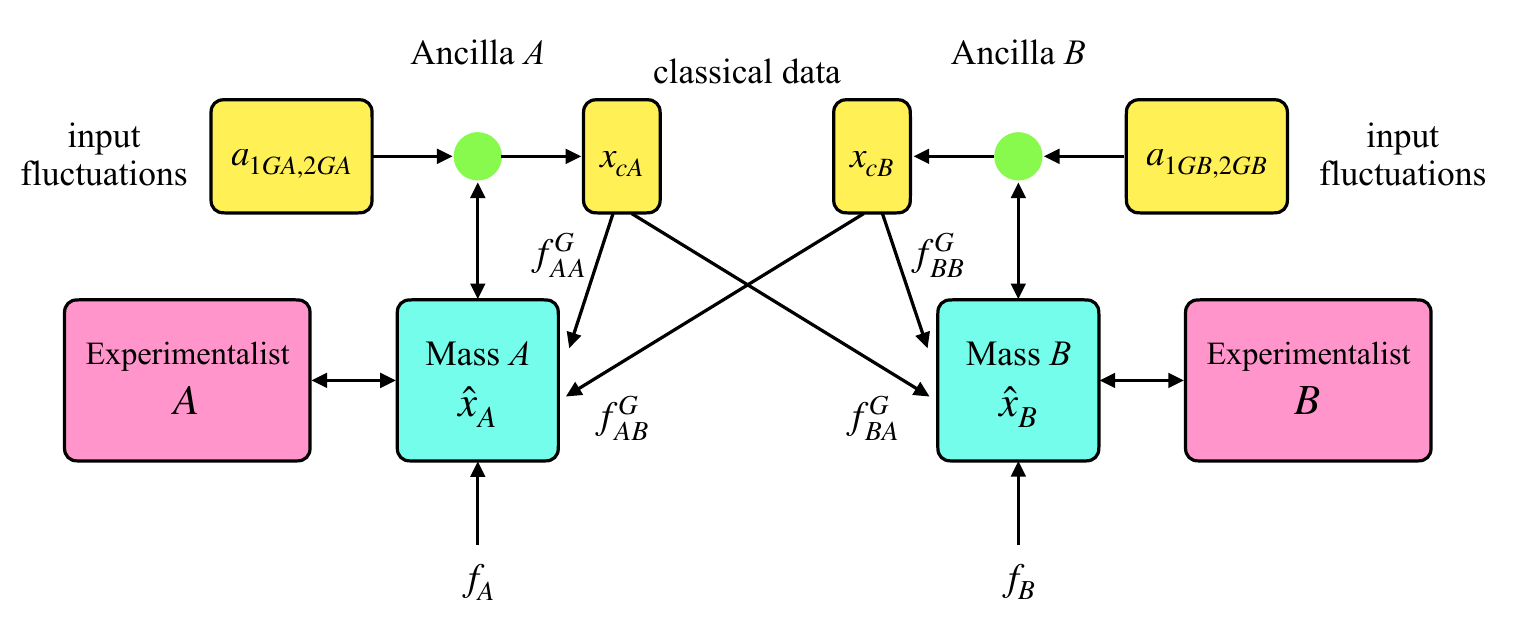}
    \caption{Schematic diagram of a continuous-monitoring and feedback realization of a linear gravitational interaction.  Masses $A$ and $B$ are each coupled to an auxiliary observer (ancilla), which measures the position of the mass and creates records $x_{cA}$ and $x_{cB}$.  The records are then fed back through the chosen gravitational response, producing self and mutual gravity.  The experimentalists can independently measure the positions of the masses; the response of the outgoing experimental fields to external forces and to fluctuations associated with the experimental measurements is the same as in the target linear theory, while the auxiliary sector contributes additional noise and correlations.  In Regimes~I and II the ancillas are local sensors; in Regime~III their inputs may be quantum-correlated, so that the auxiliary sector acts as a nonlocal mediator and can reproduce linearized quantum-gravity correlations.\label{fig:overview}}
\end{figure*}

We now make these ingredients explicit, and the overall monitoring--feedback architecture is summarized in Fig.~\ref{fig:overview}.  An auxiliary observer (an {\it ancilla}) is assigned to each massive object.  Each ancilla is driven by an input quantum field in a specified state, interacts continuously with its assigned object, and has an outgoing field that is measured continuously~\cite{braginsky1995quantum}.  The measurement record extracted from that output is a commuting classical record, and a prescribed feedback rule uses that record to generate the self/mutual-gravity forces. Different from many previous literatures that formulate the problem in the Schr\"odinger picture, the model is formulated in the Heisenberg picture, so that the input fields, output fields, feedback forces, and experimentally accessible readouts remain explicitly treated throughout the calculation.

We restrict the dynamics to be linear, but the auxiliary input state need not be vacuum or Gaussian.  Local squeezing, intra-ancilla sensing--back-action correlations, cross-object correlations, non-Gaussian states, finite propagation delays, and the tensorial response relevant to linearized gravity can all be included.  The feedback kernel fixes the desired deterministic response, while the input state fixes the additional noise and its correlations; these two parts of the prediction can therefore be characterized separately, subject to their consistency relations.  

\subsection{Correlations, regimes, and the quantum-mediated limit}

The central distinction among the models considered in the literature is the correlation structure of the auxiliary inputs.  If the ancillas associated with different objects are independent and only their commuting records are communicated for feedback, the construction is a local classical-record channel.  If the input fields share nonlocal quantum correlations, the auxiliary observer can instead act as a common mediator.  Carney \emph{et al.}~\cite{carney2025quantum} considered a related nonlocal entropic-force construction in which an auxiliary bath mediates gravity rather than serving as two independent local sensors.
The resulting classification is:
\begin{itemize}[leftmargin=*]
\item[(I)] \emph{Uncorrelated sensing and back action} --- Each mass is monitored by an independent ancilla. Each ancilla is driven independently by an input field that leads to no correlation between sensing and back-action noise quadratures.
\item[(II)] \emph{Locally correlated, cross-object independent (local-record/LOCC)} --- correlations are allowed between the sensing and back-action quadratures of each ancilla, but the different ancillas (associated with different objects) remain independent. Classical communication of the records is still used to apply the mutual-gravity feedback force.
\item[(III)] \emph{Globally correlated / common nonlocal mediator} --- the input fields of different ancillas share nonlocal quantum correlations. This regime lies outside the local-operations-and-classical-communication class.
\end{itemize}

Each regime has a distinct experimental signature, and these three regimes form a coherent hierarchy for testing the quantum nature of gravity. Specifically, in Regimes~I and II, the feedback signals are classical, the cross-object ancillas are independent, and their input states are uncorrelated, so the channel is LOCC even when each ancilla is internally squeezed or correlated. We call gravity modes in these regimes the \emph{local-record monitoring--feedback models}\,(abbreviated as the \emph{local record model}). We use “local-record” to specify the operational resource structure, and when such records are used to source gravity, the resulting a classical gravity channel. In Regime~III, the force combinations generated by the feedback commute in the instantaneous Newtonian limit; \emph{nonlocal quantum correlations} can therefore squeeze them jointly and transfer entanglement to the masses.  For a causal retarded response beyond the Newtonian regime,
since the monitoring-feedback forces have the same commutators as the quantized gravity field,  the joint input state can further be chosen such that the symmetrized correlation functions also coincide with those of any desired quantized field.
We use the term linearized quantum gravity in this operational sense: equivalence of the (retarded) linear response and correlation functions of the gravitational field, including fluctuations, without claiming to reproduce the nonlinear or ultraviolet theory.

\section{Local record model of a self-gravitating object}
\label{sec:single}

Let us consider a single object under two continuous linear measurement processes.  The first, with strength $\alpha_G$, is the auxiliary observer whose record will be used to generate gravity; the second, with strength $\alpha_E$, is the experimentalist's readout.  We denote the incoming quadratures by $a_{1G,2G}$ and $a_{1E,2E}$, respectively, with commutators
\begin{equation}
    [ a_{1E}(t),a_{2E}(t')]=[ a_{1G}(t),a_{2G}(t')]=i\delta(t-t')
\end{equation}
Throughout this paper we use dimensionless input quadratures normalized so that vacuum has single-sided symmetrized spectra
\begin{equation}
S_{a_1a_1}^{\rm vac}=S_{a_2a_2}^{\rm vac}=1,\qquad S_{a_1a_2}^{\rm vac}=0.
\end{equation}
The corresponding local uncertainty relation is
\begin{equation}
S_{a_1a_1}\,S_{a_2a_2}-|S_{a_1a_2}|^2 \;\ge\; 1,
\label{eq:spectral-convention}
\end{equation}

\begin{figure}
    \includegraphics[width=0.25\textwidth]{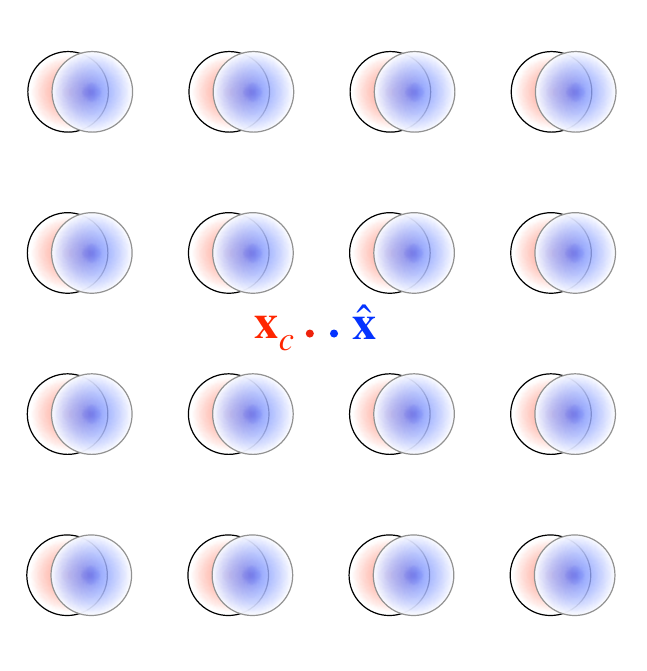}
    \caption{Illustration of a macroscopic object moving in its own classical gravity field. The classical record {$\mathbf{x}_c$} for the {center}-of-mass location of the mass creates a gravity field according to matter density $\rho_0(\mathbf{x}-\mathbf{x}_c)$, and the quantum mass distribution $\rho_0(\mathbf{x}-\hat{\mathbf{x}})$ of the object interacts with that field.  Note that the distribution $\rho_c$ is part of the classical record. The resulting self-gravity potential has a second derivative at $\hat{\mathbf{x}}=\mathbf{x}_c$ that gives rise to the Schr\"odinger--Newton frequency $\omega_{\rm SN}$, which is set by the microscopic mass distribution of the object and can be much larger than the mutual-gravity scale relevant for two-body entanglement proposals.\label{fig:self}}
\end{figure}

\subsection{Heisenberg equations and the SN frequency}
\label{subsec:SN}

In the situation of one object, we write a set of Heisenberg equations:
\begin{align}
D_m x  & = \alpha_G a_{1G} + \alpha_E a_{1E}+ f - M\omega_{\rm SN}^2 ( x- x_c),
\label{eq:grav:x} \\
b_{1G} & = a_{1G},\;\;  b_{2G}  = a_{2G} +\alpha_G x/\hbar,\;\;
x_c  =\hbar\, b_{2G}/\alpha_G, \label{eq:grav:sens}\\
b_{1E} & = a_{1E},\;\; b_{2E} = a_{2E} +\alpha_E x/\hbar.
\label{eq:grav:exp}
\end{align}
The mass evolves under the back action of both measurements, an external classical force $f$, and a self-gravity feedback term $-M\omega_{\rm SN}^2(x - x_c)$ centred at the classical record $x_c$.  The operator $D_m$ is
\begin{equation}
D_m = M[\partial_t^2  + \gamma_m \partial_t +\omega_m^2].
\end{equation}
The Schr\"odinger--Newton frequency $\omega_{\rm SN}$ is given by
\begin{equation}
\omega_{\rm SN}^2 = {Gm}/{(6\sqrt{\pi}\,x_{\rm zp}^3)},
\label{eq:omegaSN}
\end{equation}
where $m$ is the mass of each nucleus and $x_{\rm zp}$ is the one-dimensional rms localization width of nuclei near their lattice sites, typically $x_{\rm zp}\sim 10^{-12}\,$m, much smaller than the lattice spacing $\sim$~\AA.  Equation~\eqref{eq:grav:x} is the linearization of the full nonlinear self-gravity around the classical record; it is valid when the quantum uncertainty of the centre-of-mass position is much less than $x_{\rm zp}$.
The term $-M\omega_{\rm SN}^2(x-x_c)$ in \eq{eq:grav:x} has been derived before in the context of Schr\"odinger--Newton theory~\cite{yang2013macroscopic,giulini2014centre}, in which classical information for gravity is provided by the experimentalist's measurement~\cite{liu2023semiclassical,miki2025role}. However, it was not incorporated in previous models in which classical gravity was generated from auxiliary observers.  We note that it is not a classical self-force of a rigid body on itself; rather, it is associated with the second derivative of the gravitational potential energy of the mass in the classical gravity field sourced from the classical record of the mass' position.  We caution that in this model, the {\it internal matter distribution of the object has tacitly been regarded as part of the classical record}.

Let $\hat{\mathbf{x}}$ denote the center-of-mass displacement of a rigid body, and let $\varrho(\mathbf{r})$ denote its internal mass distribution in the body frame.  In a semiclassical Newtonian model, the gravitational potential is sourced by a classical record $\mathbf{x}_c$ of the body's position; the potential energy of the quantum mass (centered at $\hat{\mathbf{x}}$) in this classical field (sourced by mass centered at $\mathbf{x}_c$) is given by
\begin{equation}
    \hat V = V_{\rm self}(\hat{\mathbf{x}}, \mathbf{x}_c)
\end{equation}
with $V_{\rm self}$ defined by
\begin{equation}
V_{\rm self}(\mathbf{x} ,\mathbf{y} ) = -G\!\int d^3\mathbf{z}\, d^3\mathbf{z}
'\,\frac{\varrho(\mathbf{z}-{\mathbf{x}})\,\varrho(\mathbf{z}'-\mathbf{y})}{|\mathbf{z}-\mathbf{z}'|}\,.
\label{eq:self-energy}
\end{equation}
The situation is illustrated by Fig.~\ref{fig:self}.  The function $V_{\rm self}$ depends only on $\mathbf{x}-\mathbf{y}$ and its first {derivative} at $\mathbf{x}=\mathbf{y}$ vanishes by symmetry.  For small values of $|\mathbf{x}-\mathbf{y}|$, assuming isotropy of mass distribution near lattice sites, we obtain
\begin{equation}
V_{\rm self} (\mathbf{x},\mathbf{y}) = V_0 + \frac{1}{2} M\omega_{\rm SN}^2 |\mathbf{x}-\mathbf{y}|^2 + O(|\mathbf{x}-\mathbf{y}|^4).
    \label{eq:Kdef}
\end{equation}
For a solid in which nuclei of mass $m$ are localized near lattice sites with Gaussian width $x_{\rm zp}$, $\omega_{\rm SN}$ is given by Eq.~\eqref{eq:omegaSN}.  The important point is that the cutoff is not the macroscopic size of the body, but the microscopic localization of nuclear mass at lattice sites; for a crystal $\omega_{\rm SN}$ is roughly two orders of magnitude larger than the homogeneous-density estimate~\cite{yang2013macroscopic,helou2017measurable}, and is the same scale that controls the Causal-Conditional formulation of SN theory~\cite{liu2023semiclassical,liu2025semiclassical,miki2025role}.

\subsection{Classical-Gravity Noise (CGN) spectrum from the monitoring--feedback process}

In Eq.~\eqref{eq:grav:sens}, let us comment that
the classical record for the mass' center-of-mass position is equated to the Heisenberg operator of the output quadrature, $x_c = \hbar\,b_{2G}/\alpha_G$.   It is important to recognize that, since all the observables we measure are outgoing fields (from the auxiliary observers) that have commuting Heisenberg operators, {\it the principle of deferred measurements} allows us to treat $b_{2G}$ and $x_c$ as Heisenberg operators within the Heisenberg equations and perform state reduction only at the end of the calculation.

Physically, the term $\alpha_G a_{1G}$ in \eq{eq:grav:x} is the inevitable back action noise imposed by the measurement process.  From \eq{eq:grav:sens}, $x - x_c = -\hbar\, a_{2G}/\alpha_G$, so the self-gravity feedback contributes an imprecision noise of $+(\hbar M\omega_{\rm SN}^2/\alpha_G)\,a_{2G}$, and the equation of motion becomes
\begin{equation}
D_m x = \alpha_E a_{1E} + f + \underbrace{\left[\alpha_G a_{1G} + ({\hbar M\omega_{\rm SN}^2}/{\alpha_G})\,a_{2G}\right]}_{\mathscr{F}}.
\end{equation}
We identify the bracketed combination as the classical-gravity noise force on the object,
\begin{equation}
\mathscr{F} = \alpha_G a_{1G} + [{\hbar M\omega_{\rm SN}^2}/{\alpha_G}]\,a_{2G}.
\label{eq:fg}
\end{equation}
with first and second terms corresponding to back-action and imprecision noise, respectively. From the point of view of the experimentalist who have access to $b_{1,2E}$, the only effect from monitoring-feedback gravity is the additional classical-gravity noise force $\mathscr{F}$, and now let us calculate its spectrum.

For ancilla in vacuum, using the convention of \eq{eq:spectral-convention},
\begin{equation}
S_{\mathscr{F}} = \alpha_G^2 + [\hbar^2 M^2\omega_{\rm SN}^4/{\alpha_G^2}] \;\ge\; 2\hbar M\omega_{\rm SN}^2,
\label{eq:Sfg-min}
\end{equation}
with the inequality saturated at $\alpha_G^2 = \hbar M\omega_{\rm SN}^2$ by the arithmetic--geometric-mean inequality.  This bound holds for any monitoring scheme with uncorrelated sensing and back-action quadratures.  Its minimum is at the same order as noise modifications predicted by Schr\"odinger--Newton theories~\cite{miki2025role,liu2023semiclassical,liu2025semiclassical}. Note that the $S_{\mathscr{F}}$ here is equal to the {\it free-mass Standard  Quantum Limit} for force sensing, at $\omega_{\rm SN}$; for a free mass, the position and momentum uncertainty this force produces near $\omega_{\rm SN}$ are comparable to zero-point position and momentum fluctuations of a harmonic oscillator at $\omega_{\rm SN}$. This is one of the central results of the paper.

If, instead, the ancilla input quadratures are correlated, the spectrum reads
\begin{equation}
S_{\mathscr{F}} = \alpha_G^2 S_{a_1a_1}^{AG} + [{\hbar^2 M^2\omega_{\rm SN}^4}/{\alpha_G^2}]\,S_{a_2a_2}^{AG} + 2\hbar M\omega_{\rm SN}^2\,\mathrm{Re}\,S_{a_1a_2}^{AG},
\end{equation}
with the cross-spectral density matrix constrained by \eq{eq:spectral-convention}.  An appropriately squeezed input can reduce the force noise below the uncorrelated bound --- even to zero --- at the cost of indefinitely anti-squeezing its conjugate quadrature.  Equation~\eqref{eq:Sfg-min} should therefore be read as the noise floor of the simplest local classical-record construction with uncorrelated sensing and back action; allowing squeezed auxiliary inputs is a quantum resource that relaxes this floor.   The discussion here is analogous to the circumvention of the Standard Quantum Limit in linear quantum measurement processes by injecting the appropriate squeezed vacuum, as has been discussed in Refs.~\cite{kimble2001conversion} and implemented in gravitational-wave detectors~\cite{jia2024squeezing}. We note that the existence of the two types of classical-gravity noise, as well as their trade-off, has been discussed by Kafri, Taylor, and Milburn~\cite{kafri2014classical}, as well as by Oppenheim and collaborators~\cite{oppenheim2023gravitationally}. 

Even though the additional classical-gravity {noise} can be suppressed {indefinitely} for a single mass, the {anti-squeezing} of the conjugate quadrature will have consequences when we consider the mutual gravity of two masses in Sec.~\ref{sec:twoobjects}.  In this section,  we find it convenient, with correlations allowed, to define reorganized input quadratures that absorb the self-gravity coupling,
\begin{equation}
u_1 = a_{1G} + [{\hbar M\omega_{\rm SN}^2}/{\alpha_G^2}]\,a_{2G},\qquad u_2 = a_{2G},
\end{equation}
so that $\mathscr{F} = \alpha_G u_1$ and $S_{\mathscr{F}} = \alpha_G^2 S_{u_1u_1}^G$.  We can then introduce a characteristic classical-gravity frequency $\omega_{\rm CG}$ via
\begin{equation}
S_{\mathscr{F}} \equiv 2\hbar M\omega_{\rm CG}^2\,,\quad \omega_{\rm CG}^2 = \alpha_G^2 S_{u_1u_1}^G/(2\hbar M).
\end{equation}
This provides a basis for discussing the level of the additional classical-gravity noise for a single object. Numerically, it has the same value as the free-mass Standard Quantum Limit at frequency $\omega_{\rm CG}$.  If the ancilla has vacuum input fields, we have  $\omega_{\rm CG}\ge\omega_{\rm SN}$; with correlations allowed between quadratures and squeezing of $u_1$, $\omega_{\rm CG}$ can {\it in principle approach zero}, when we choose the input state as an infinitely squeezed state with squeezing angle $\pi/4$.  Such correlations and squeezing can be viewed as quantum resources required to allow the single object to have a reduced level of additional classical-gravity noise.

\section{Mutual gravity between two objects generated by local record}
\label{sec:twoobjects}

We now add a second body while retaining the local-record architecture: each object is monitored by its own ancilla, and only the commuting records are allowed to be shared between the bodies and used for feedback to generate gravity.  This section therefore covers two cases introduced in Sec.~\ref{sec:modeloverview}: Regime~I, with uncorrelated sensing and back-action, and Regime~II, with arbitrary correlations within each ancilla but no cross-object input correlations.  The distinction is important.  In Regime~I, the two-body force noise is dominated by the single-object self-gravity scale.  In Regime~II, local squeezing can redistribute or reduce the noise seen by either object separately, but the joint force-noise matrix remains constrained by the local-channel determinant bound and the interaction remains non-entangling.  Only in Sec.~\ref{sec:RIII} do we allow cross-object quantum correlations.  We restrict the present section to instantaneous Newtonian mutual gravity; the retarded case is treated later.

\subsection{Equations of motion}

Here we introduce classical mutual gravity through feedback based on the classical position records. These records are obtained by the two auxiliary monitors and used to apply feedback forces to the objects. The corresponding mutual gravity potential, realized via feedback, is
\begin{align}
\label{eq:Vmutual}
    V_{\rm mutual}=-\frac{2GM_AM_B}{d^3}\left[\hat x_A(x_{cA}-x_{cB})+\hat x_B(x_{cB}-x_{cA})\right]
\end{align}
Similar to the single-object case, we write down the Heisenberg equations for the two-body system,
\begin{align}
D_A x_A &= \alpha_{AG} a_{1AG} + \alpha_{AE} a_{1AE} + f_A \nonumber\\
&\quad - K_{AA}(x_A - x_{cA}) - K_{AB}(x_{cB} - x_{cA}),\\
D_B x_B &= \alpha_{BG} a_{1BG} + \alpha_{BE} a_{1BE} + f_B \nonumber\\
&\quad - K_{BB}(x_B - x_{cB}) - K_{AB}(x_{cA} - x_{cB}),
\end{align}
with self- and mutual-gravity ``spring constants'' given by
\begin{equation}
K_{AA} = M_A\omega_{\rm SNA}^2,\quad K_{BB} = M_B\omega_{\rm SNB}^2,
\end{equation}
and
\begin{equation}
K_{AB} = \sqrt{M_A M_B}\,\omega_G^2 = {2GM_AM_B}/{d^3}.
\end{equation}

Here  $\omega_G$ is the time scale of gravitational interaction, which is typically maximized when the mases are comparable, in which case $\omega_G^2 = 2GM/d^3$.  Note that self gravity is much stronger than mutual gravity, and $K_{AA}, K_{BB} \gg K_{AB}$.  For silicon test masses, $\omega_{\rm SN}\sim 0.036\,\mathrm{s}^{-1}$, while at centimetre-scale separations $\omega_G \sim \sqrt{2GM/d^3}\sim\sqrt{\pi G\rho/3} \approx 4\times 10^{-4}\,\mathrm{s}^{-1}$, so $K_{AA}/K_{AB} \sim (\omega_{\rm SN}/\omega_G)^2\sim 10^4$.

For the auxiliary monitors, we similarly have 
\begin{align}
b_{1AG} &= a_{1AG},\; b_{2AG} = a_{2AG} + \frac{\alpha_{AG} x_A}{\hbar},\; x_{cA} = \frac{\hbar b_{2AG}}{\alpha_{AG}},\\
b_{1BG} &= a_{1BG},\; b_{2BG} = a_{2BG} + \frac{\alpha_{BG} x_B}{\hbar},\; x_{cB} = \frac{\hbar b_{2BG}}{\alpha_{BG}},
\end{align}
and for the experimental readouts,
\begin{align}
b_{1AE} &= a_{1AE},\quad b_{2AE} = a_{2AE} + \alpha_{AE} x_A/\hbar,\\
b_{1BE} &= a_{1BE},\quad b_{2BE} = a_{2BE} + \alpha_{BE} x_B/\hbar.
\end{align}
The experimental outputs depend on the mechanical responses $x_A, x_B$, which obey the usual Newtonian equations together with additional measurement-induced forces.  

In the Heisenberg equations for $x_A$ and $x_B$, we can separate  the coherent Newtonian response from the additional monitoring--feedback noise, writing
\begin{align}
D_A x_A &= \alpha_{AE} a_{1AE} + f_A - K_{AB}(x_B - x_A) + \mathscr{F}_A,\\
D_B x_B &= \alpha_{BE} a_{1BE} + f_B - K_{AB}(x_A - x_B) + \mathscr{F}_{B},
\end{align}
where the coherent mutual-gravity term $-K_{AB}(x_B - x_A)$ matches the standard quantized Newtonian interaction at linear input--output level, and $\mathscr{F}_A, \mathscr{F}_B$ collect the additional noises:
\begin{align}
\mathscr{F}_A &= \left[\alpha_{AG} a_{1AG} + \frac{(K_{AA}+K_{AB})\hbar}{\alpha_{AG}}\,a_{2AG}\right] - \frac{\hbar K_{AB}}{\alpha_{BG}}\,a_{2BG},\label{eq:cqa}\\
\mathscr{F}_B &= \left[\alpha_{BG} a_{1BG} + \frac{(K_{BB}+K_{AB})\hbar}{\alpha_{BG}}\,a_{2BG}\right] - \frac{\hbar K_{AB}}{\alpha_{AG}}\,a_{2AG}. \label{eq:cqb}
\end{align}

An alternative construction of the mutual-gravity interaction is to let the gravitational potential sourced by the classical record of one object act on the other. This is implemented by the following interaction Hamiltonian
\begin{align}
    V_{\rm mutual}^\prime=-\frac{GM_AM_B}{d^3}\left[(\hat x_A-x_{cB})^2+(\hat x_B-x_{cA})^2\right]
\end{align}
which differs from Eq.~\eqref{eq:Vmutual}.
Thus, the potential determined by the record of object $B$ acts on object $A$, and vice versa. This construction gives the same coherent Newtonian terms as those in Eqs.~\eqref{eq:cqa} and \eqref{eq:cqb}, while the additional force noises are replaced by
\begin{align}
\label{eq:fap}
    \mathscr{F}_A^{\prime}
    &=
    \left[
        \alpha_{AG}a_{1AG}
        +\frac{\hbar K_{AA}}{\alpha_{AG}}a_{2AG}
    \right]
    -\frac{\hbar K_{AB}}{\alpha_{BG}}a_{2BG},\\
    \label{eq:fbp}
    \mathscr{F}_B^{\prime}
    &=
    \left[
        \alpha_{BG}a_{1BG}
        +\frac{\hbar K_{BB}}{\alpha_{BG}}a_{2BG}
    \right]
    -\frac{\hbar K_{AB}}{\alpha_{AG}}a_{2AG}.
\end{align}
Compared with the original construction, this alternative removes noise terms $\hbar K_{AB}a_{2AG}/\alpha_{AG}$ within $\mathcal{F}_A$ and
$\hbar K_{AB}a_{2BG}/\alpha_{BG}$ within $\mathcal{F}_B$. These are local input noise fluctuations imposed on the same local object, but with amplitude determined by the mutual coupling constant $K_{AB}$.  Consequently, when
$K_{AA},K_{BB}\gg K_{AB}$ and the self-gravity contributions are not canceled by specially chosen local correlations, the two constructions give the same leading-order force noise.  Furthermore, as we allow local correlations, i.e., between $a_{1AG}$ and $a_{2AG}$, and between $a_{1BG}$ and $a_{2BG}$, the two formulations are related by altering local correlation structures, see Sec.~\ref{subsubsec:iib}.

In the following, we adopt the former construction, which more straightforwardly leads to the negative correlations between mutual-gravity force noises found in Ref.~\cite{kryhin2025distinguishable}.

\subsection{Ancilla in vacuum: Regime~I}

As in the single object case, the simplest construction is to have vacuum input fields for the ancillas.   We refer to it as Regime~I.  The spectral densities are
\begin{align}
S_{\mathscr{F}_A} &= \alpha_{AG}^2 + \frac{\hbar^2(K_{AA}+K_{AB})^2}{\alpha_{AG}^2} + \frac{\hbar^2 K_{AB}^2}{\alpha_{BG}^2},\\
S_{\mathscr{F}_B} &= \alpha_{BG}^2 + \frac{\hbar^2(K_{BB}+K_{AB})^2}{\alpha_{BG}^2} + \frac{\hbar^2 K_{AB}^2}{\alpha_{AG}^2},
\end{align}
with cross-spectrum
\begin{equation}
S_{\mathscr{F}_A\mathscr{F}_B} = -\hbar^2 K_{AB}\!\left[\frac{K_{AA}+K_{AB}}{\alpha_{AG}^2} + \frac{K_{BB}+K_{AB}}{\alpha_{BG}^2}\right].
\end{equation}
Although the sensing channels of $A$ and $B$ are independent, $\mathscr{F}_A$ and $\mathscr{F}_B$ are anti-correlated because both involve the differential mutual-gravity force.  This is consistent with the findings of Ref.~\cite{kryhin2025distinguishable}.  

Since $\omega_{\rm SN}\gg\omega_G$, the optimal $\alpha_{AG}, \alpha_{BG}$ are essentially the single-object values, and the additional companion-induced corrections are suppressed by $\omega_G^2/\omega_{\rm SN}^2$:
\begin{align}
S_{\mathscr{F}_{A}} &\ge 2\hbar M_A\omega_{\rm SNA}^2\bigl[1 + O(\omega_G^2/\omega_{\rm SNA}^2)\bigr],\\
S_{\mathscr{F}_{B}} &\ge 2\hbar M_B\omega_{\rm SNB}^2\bigl[1 + O(\omega_G^2/\omega_{\rm SNB}^2)\bigr].
\end{align}

In other words, in Regime I, with vacuum input fields at each ancilla, each object's force noise is controlled primarily by its own self-gravity additional noise {$2\hbar M\omega_{\mathrm{SN}}^2$}, with mutual-gravity corrections suppressed by $\omega_G^2/\omega_{\rm SN}^2$.   This is a very important point: testing the self-gravity of a single object is an important stepping stone toward testing mutual gravity and gravity-induced entanglement between two objects.  We return to this experimental hierarchy in Sec.~\ref{sec:experiments}.

\subsection{Independent ancillas with local correlations: Regime II}

We now gradually allow more correlations between auxiliary input fields, distinguishing between correlations between local quadratures (within a single ancilla) and correlations across the ancillas of the two objects. Once such correlations are allowed, the same physical auxiliary-noise realization can be represented in different local quadrature bases, provided that the auxiliary state is transformed accordingly. Since the choice of basis and the corresponding state description are model dependent, we formulate the gravity noise below in convenient bases that makes the characteristic structure of each subregime transparent.

\subsubsection{Regime~II: local sensing with local correlations}
\label{sec:RII}

Regime~II is the class in which the sensing of objects $A$ and $B$ remains local: classical information about $x_{cA}$ and $x_{cB}$ may be communicated across in order to apply the Newtonian feedback force {(as it must, in order to compute the inter-body force)}, but the auxiliary input fields that obtain the two records are not themselves a common sensing system.  This is the local-sensing or LOCC class.  It can produce correlated forces, but it cannot produce gravity-induced entanglement.

To allow the local cancellation of the large self-gravity noise, we first absorb the self-gravity part into locally re-organized quadratures (the single-object reorganization introduced in Eq.~\eqref{eq:fg}):,
\begin{align}
u_{1AG} &= a_{1AG} + [{\hbar K_{AA}}/{\alpha_{AG}^2}]\,a_{2AG},\quad u_{2AG} = a_{2AG},\\
u_{1BG} &= a_{1BG} + [{\hbar K_{BB}}/{\alpha_{BG}^2}]\,a_{2BG},\quad u_{2BG} = a_{2BG}.
\end{align}
and then  write
\begin{align}
\mathscr{F}_A &= \alpha_{AG} u_{1AG} + [{\hbar K_{AB}}/{\alpha_{AG}}]\,u_{2AG} - [{\hbar K_{AB}}/{\alpha_{BG}}]\,u_{2BG},\\
\mathscr{F}_B &= \alpha_{BG} u_{1BG} + [{\hbar K_{AB}}/{\alpha_{BG}}]\,u_{2BG} - [{\hbar K_{AB}}/{\alpha_{AG}}]\,u_{2AG}.
\end{align}
In this form the large self-gravity constants $K_{AA}, K_{BB}$ have been absorbed into local correlations between sensing and back-action quadratures of each ancilla.  Only the mutual-gravity coupling $K_{AB}$ and the local monitoring strengths remain.  In particular, when $K_{AB} = 0$, the noise on each object is controlled by $u_{1AG}$ or $u_{1BG}$ alone, which can be made arbitrarily small by local squeezing --- recovering the single-object discussion.  Once mutual gravity is present, however, local sensing leaves a residual constraint, which we now examine in two sub-regimes.

\subsubsection{Regime~IIa: restricted local correlations}

Within Regime~II, we consider first the restricted case in which $u_{1\cdot G}, u_{2\cdot G}$ [the above newly defined quadratures directly proportional to the total single-mass force noise] are vacuum, and $\alpha_{AG}, \alpha_{BG}$ scale with mass through a {\it fundamental classical-gravity frequency scale} $\omega_{\rm CG}$,
\begin{equation}
\label{eqdefcg}
    \alpha_{AG}^2 = 2\hbar M_A\omega_{\rm CG}^2,\qquad \alpha_{BG}^2 = 2\hbar M_B\omega_{\rm CG}^2.
\end{equation}
In absence of a second object, or when $A$ and $B$ are infinitely far away, classical-gravity force noises on $A$ and $B$ are given by 
\begin{align}
\label{regimeiisingle}
S_{\mathscr{F}_A}^{\rm single} = 2\hbar M_A\omega_{\rm CG}^2,\quad 
S_{\mathscr{F}_B}^{\rm single} = 2\hbar M_B\omega_{\rm CG}^2\,.
\end{align}
Note that, by allowing local correlations of input fields,  $\omega_{\rm CG}$ is a free parameter of the model. Single-object experiments can put bounds on $\omega_{\rm CG}$, but it does not have a natural scale. 

Now, as $A$ and $B$ are placed in the vicinity of each other, the classical-gravity force  spectral densities are given by \begin{align}
\label{regimeiiself}
S_{\mathscr{F}_A} &= \hbar\bigl[2M_A\omega_{\rm CG}^2 + (M_A+M_B)\omega_G^4/2\omega_{\rm CG}^2\bigr],\\
S_{\mathscr{F}_B} &= \hbar\bigl[2M_B\omega_{\rm CG}^2 + (M_A+M_B)\omega_G^4/2\omega_{\rm CG}^2\bigr],
\end{align}
with cross spectrum 
\begin{equation}
\label{regimeiiacorr}
    S_{\mathscr{F}_A\mathscr{F}_B} = -\hbar(M_A+M_B)\omega_G^4/2\omega_{\rm CG}^2
    \end{equation}
Minimizing each per-object spectrum over $\omega_{\rm CG}$  gives
\begin{align}
S_{\mathscr{F}_A} &\ge 2\hbar\sqrt{M_A(M_A+M_B)}\,\omega_G^2,\\
S_{\mathscr{F}_B} &\ge 2\hbar\sqrt{M_B(M_A+M_B)}\,\omega_G^2,
\end{align}
saturated at $\omega_{\rm CG}^4 = (1+M_B/M_A)\omega_G^4$ for $A$ and $\omega_{\rm CG}^4 = (1+M_A/M_B)\omega_G^4$ for $B$. For comparable masses the floor is $\sim 2\sqrt{2}\,\hbar M\omega_G^2$.    

We note that, even though single-object noise due to sensing and feedback can be made arbitrarily small by reducing $\omega_{\rm CG}$, the presence of a second object causes an additional noise that blows up when $\omega_{\rm CG} \ll \omega_G$.  The trade-off prevents the establishment of quantum entanglement via gravity, as they stay within the bound obtained by Ref.~\cite{tilloy2016sourcing,tilloy2017principle,fabiano2026minimal}.

Similar to Regime I, we once more find a negative cross-correlation~\eqref{regimeiiacorr} between $\mathscr{F}_A$ and $\mathscr{F}_B$, due to the differential nature of the mutual-gravity force, consistent with Ref.~\cite{kryhin2025distinguishable}.
 However, the presence of this correlation does not allow the channel to become entangling, as our channel is explicitly LOCC.

\begin{table*}
    \centering
    \footnotesize
    \begin{tabular}{c|c|c|c|c}
         Regime & \begin{tabular}{c}Local  \\Correlations \end{tabular}
        & \begin{tabular}{c}Cross-Object \\ Sensing\end{tabular} & Single-mass effect & Two-mass effect \\
        \hline\hline
         I & None & None &  $S_{f_G}\ge 2\hbar M\omega_{\rm SN}^2$ [Eqs.~\eqref{eq:Kdef} and \eqref{eq:Sfg-min}] & SN-dominated; mutual corrections $O(\omega_G^2/\omega_{\rm SN}^2)$\\
         \hline
         IIa & Restricted & None & $S_{f_G}=2\hbar M\omega_{\rm CG}^2$ [Eq.~\eqref{eqdefcg} and \eqref{regimeiiself}] & \begin{tabular}{c} Presence of second object causes  noise [Eqs.~\eqref{regimeiiself}--\eqref{regimeiiacorr}] \\ floor lower-limited by $\stackrel{>}{_\sim} 2\sqrt{2}\,\hbar M\omega_G^2$ for each; no GIE \end{tabular}\\
        IIb & General & None & Unconstrained & Eqs.~\eqref{regimeiibaa}--\eqref{regimeiibab}; Joint determinant bound [Eq.~\eqref{eq:det-bound-main}]; no GIE \\
         \hline
        III & General & Allowed & Unconstrained &  \begin{tabular}{c}
        GIE allowed; Linearized Quantum Gravity realizable; \\
        Quantum fluctuations of the gravity field
        \end{tabular}
         \\
\end{tabular}
    \caption{Classification of auxiliary correlation structures.  As the allowed correlations are progressively relaxed from Regime~I to Regime~III, the additional noise can be reduced.  Regime~II is the local-classical-channel/LOCC class containing the Kafri--Taylor and Fabiano \emph{et al.} bounds; Regime~III is the nonlocal-mediator class exemplified by the entropic-force model of Carney \emph{et al.}, in which gravity-induced entanglement is possible and the discriminant from perturbative gravitons is the noise/correlation structure, although Regime III can include linearized quantum gravity for linear systems.  \emph{The key distinction between Regimes~II and III is not whether the resulting forces are correlated --- correlated forces already occur in Regime~II --- but whether the underlying sensing records remain local.}}
    \label{tab:correlations}
\end{table*}

\subsubsection{Regime~IIb: general local correlations}
\label{subsubsec:iib}

For Regime~IIb we allow an arbitrary correlation structure inside each local auxiliary channel, but still no auxiliary correlations between $A$ and $B$.  It is useful to absorb both the self-gravity term and the local part of the mutual feedback into
\begin{align}
v_{1AG} &= a_{1AG} + \frac{\hbar(K_{AA}+K_{AB})}{\alpha_{AG}^2}\,a_{2AG},\quad v_{2AG} = a_{2AG},\\
v_{1BG} &= a_{1BG} + \frac{\hbar(K_{BB}+K_{AB})}{\alpha_{BG}^2}\,a_{2BG},\quad v_{2BG} = a_{2BG}.
\end{align}
[Note that for the alternative formulation given by Eqs.~\eqref{eq:fap} and \eqref{eq:fbp}, we simply remove the $K_{AB}$ terms in the above equations.]
Then the additional forces take the simple form
\begin{align}
\mathscr{F}_{A} &= \alpha_{AG} v_{1AG} - \frac{\hbar K_{AB}}{\alpha_{BG}}\,v_{2BG},\\
\mathscr{F}_{B} &= \alpha_{BG} v_{1BG} - \frac{\hbar K_{AB}}{\alpha_{AG}}\,v_{2AG}.
\end{align}
Assuming no correlations between the $A$ and $B$ gravity channels, the symmetrized spectra are
\begin{align}
\label{regimeiibaa}
S_{\mathscr{F}_A} &= \alpha_{AG}^2 S_{v_1v_1}^{AG} + \frac{\hbar^2 K_{AB}^2}{\alpha_{BG}^2}\,S_{v_2v_2}^{BG},\\
S_{\mathscr{F}_B} &= \alpha_{BG}^2 S_{v_1v_1}^{BG} + \frac{\hbar^2 K_{AB}^2}{\alpha_{AG}^2}\,S_{v_2v_2}^{AG},\\
\label{regimeiibab}
S_{\mathscr{F}_A\mathscr{F}_B} &= -\hbar K_{AB}\bigl(S_{v_1v_2}^{AG} + S_{v_1v_2}^{BG}\bigr).
\end{align}
This makes the central point transparent.  The individual force noises are {\it not} independently bounded: one can make $S_{v_1v_1}^{AG}$ and $S_{v_2v_2}^{BG}$ small and hence suppress $S_{\mathscr{F}_A}$.  But the local uncertainty relations \eqref{eq:spectral-convention} then force the conjugate spectra $S_{v_2v_2}^{AG}$ and $S_{v_1v_1}^{BG}$ to grow, which increases the noise on the joint channel.  We also notice that $\mathscr{F}_A$ and $\mathscr{F}_B$ can be correlated.

Different local choices of quadratures are related by canonical transformations and therefore provide different parametrizations of the same class of local auxiliary sector. In particular, the $\hat v$ quadratures of Regime~IIb differ from the $\hat u$ quadratures of Regime~IIa only by a local transformation that absorbs the local part of the mutual-gravity feedback. Consequently, the vacuum state in the $\hat u$ basis corresponds to a particular locally correlated state in the $\hat v$ basis, where the spectra are given by $S_{v_1v_1}=1+(\hbar K_{AB}/\alpha_G^2)^2$, $S_{v_2v_2}=1$, and $S_{v_1v_2}=S_{v_2v_1}=\hbar K_{AB}/\alpha_G^2$. Regime~IIa can therefore be regarded as a restricted realization within the more general Regime~IIb.

To absorb the explicit monitoring-strength prefactors, we introduce a further local canonical rescaling
\begin{align}
    w_{1jG}&=
    \frac{\alpha_{jG}}{\sqrt{\hbar K_{AB}}}v_{1jG},\quad
    w_{2jG}=
    \frac{\sqrt{\hbar K_{AB}}}{\alpha_{jG}}v_{2jG}.
    \label{w-quadratures}
\end{align}
This transformation also preserves the canonical commutation relations. Since the two rescaling factors are reciprocal, the $v$ and $w$ quadratures are related by a local single-mode squeezing transformation. The force noises then take the form
\begin{align}
    \mathscr F_A&=
    \sqrt{\hbar K_{AB}}
    \left(w_{1AG}-w_{2BG}\right)\\
    \mathscr F_B&=
    \sqrt{\hbar K_{AB}}
    \left(w_{1BG}-w_{2AG}\right).
\end{align}
Thus, the monitoring strengths $\alpha_{AG}$ and $\alpha_{BG}$ do not label distinct sets of attainable force noise spectra. Their explicit appearance in the force noise can be absorbed into local rescalings of the corresponding auxiliary quadratures. Equivalently, changing the monitoring strengths changes the local spectral matrices of the $w$ quadratues.

Mathematically, for any choice of $\alpha_{AG}, \alpha_{BG}$ and any local Gaussian state of the auxiliary inputs, one has
\begin{equation}
S_{\mathscr{F}_A\mathscr{F}_A}\,S_{\mathscr{F}_B\mathscr{F}_B} - |S_{\mathscr{F}_A\mathscr{F}_B}|^2 \;\ge\; 4\hbar^2 K_{AB}^2,
\label{eq:det-bound-main}
\end{equation}
which we prove explicitly in Appendix~\ref{app:noent}.  This is the same noise structure that appears in local classical-channel gravity~\cite{kafri2014classical,fabiano2026minimal}.

\section{Experimental implications of local-record models}
\label{sec:experiments}

The preceding sections established the two local-record regimes. We now translate their predictions into experimental targets, focusing on the force noise associated with local auxiliary monitoring. Regime~I provides the single-body self-gravity target, while Regime~II provides the lower mutual-gravity scale and the joint noise constraint associated with local classical channels. These are logically distinct tests: the first concerns the absolute noise of one monitored mass, whereas the second concerns how noise is distributed between two masses and whether their interaction can generate entanglement. We defer the nonlocal Regime~III and the effects of propagation delay to Sec.~\ref{sec:RIII}. For evaluating experimental feasibility, the free-mass Standard Quantum Limit for force sensing is~\cite{braginsky1995quantum}
\begin{equation}
    S_F^{\rm SQL} (\omega) = 2\hbar M\omega^2\,.
\end{equation}

\subsection{Uncorrelated local monitoring (Regime~I): the single-body target}

Regime~I gives a minimum force noise on each individual mass,
\begin{equation}
S_{\mathscr{F}} \ge 2\hbar M\omega_{\rm SN}^2,
\label{eq:RI-bound}
\end{equation}
under the assumption of uncorrelated/vacuum auxiliary monitoring.  Two phenomenological connections sharpen the experimental positioning of this bound: to Di\'osi--Penrose collapse models, and to Schr\"odinger--Newton and Correlated World-Line theories.

\subsubsection{Di\'osi--Penrose collapse: theory and phenomenology}

A minimum force noise sourced by a continuous mass-density monitor is structurally close to gravity-related collapse models, originally introduced to produce objective wavefunction collapse~\cite{bassi2013models,bassi2023collapse}.  The most directly connected case is the Di\'osi--Penrose (DP) model, in which a Gaussian-smeared matter-density operator $\hat\varrho_{\sigma_{\rm DP}}$ is continuously monitored.  The DP master equation is
\begin{equation}
\mathcal{L}_{\rm DP}\hat\rho = -\frac{G}{2\hbar}\!\iint\!\frac{d^3\mathbf{s}_1\, d^3\mathbf{s}_2}{|\mathbf{s}_1-\mathbf{s}_2| }\bigl[\hat\varrho_{\sigma_{\rm DP}}(\mathbf{s}_1),[\hat\varrho_{\sigma_{\rm DP}}(\mathbf{s}_2), \hat\rho]\bigr].
\end{equation}
The DP master equation can be viewed as a continuous monitoring of the mass density operator $\hat\varrho_{\sigma_{\rm DP}}$ with a kernel given by the Newtonian Green function~\cite{diosi1987universal,penrose1996gravity}. Here the regularization scale $\sigma_{\rm DP}$ is necessary to prevent divergence in the case of point particles; it is a free parameter that sets the spatial resolution of the mass-density monitoring, and hence the strength of the collapse. 

The same Lindbladian was obtained by Nimmrichter and Hornberger~\cite{nimmrichter2015stochastic} as a Markovian linearization of the Schr\"odinger--Newton theory, and a closely related density-monitoring Lindbladian appears in the CQ framework~\cite{oppenheim2023postquantum,oppenheim2023gravitationally,kryhin2025distinguishable,fabiano2026minimal}.  In this sense DP can be viewed as a special case of the monitoring--feedback construction of semiclassical gravity, and existing experimental tests of DP are also tests of Regime~I.

For motion of small magnitude around an equilibrium position, the DP Lindbladian linearizes to a centre-of-mass momentum diffusion.  For a solid with monatomic lattice spacing $a$~\cite{nimmrichter2015stochastic}, for {single-sided} spectrum, we have
\begin{equation}
S_{f_{\rm DP}} = \frac{G\hbar}{3\sqrt\pi}\Bigl(\frac{a}{\sigma_{\rm DP}}\Bigr)^3 M\rho \;\equiv\; 2\hbar M\omega_{\rm DP}^2,
\end{equation}
defining the DP frequency
\begin{equation}
\omega_{\rm DP} = \sqrt{\frac{G\rho}{6\sqrt\pi}}\,\Bigl(\frac{a}{\sigma_{\rm DP}}\Bigr)^{3/2}.
\end{equation}
Combining with \eq{eq:omegaSN} for a monatomic crystal (in which $a^3\rho/m = 1$) gives the parameter-free scaling
\begin{equation}
\omega_{\rm DP} = {\omega_{\rm SN}}\,\Bigl(\frac{x_{\rm zp}}{\sigma_{\rm DP}}\Bigr)^{3/2}.
\label{eq:DPscaling}
\end{equation}
We note that the original  motivation of $\sigma_{\rm DP}$ and $x_{\rm zp}$ are different: $\sigma_{\rm DP}$ is for regularization when sensing the local gravitational acceleration near point particles, while $x_{\rm zp}$ results from treating matter distribution with the massive object as a classical record, with mass concentrated within $\sim x_{\rm zp}$ around lattice sites.  Nevertheless, they play the same mathematical roles and appear the same way, {respectively}, in $\omega_{\rm DP}$ and $\omega_{\rm SN}$.   For $\sigma_{\rm DP} \ll x_{\rm zp}$ the DP regulator is finer than the physical lattice scale and DP causes higher noise than SN, while for $\sigma_{\rm DP} \gg x_{\rm zp}$, DP is coarser and predicts less noise. We shall return to a conceptual discussion of DP in {Regime II below}.

\subsubsection{Current and projected experimental bounds}

The DP regulator is bounded below by $\sigma_{\rm DP}\gtrsim 2\times 10^{-13}\,$m by LISA Pathfinder~\cite{armano2024depth,helou2017lisa,dai2024updating,carlesso2016experimental}, with comparable bounds from micro-cantilever experiments~\cite{vinante2016upper}.   Advanced LIGO operates close to the free-mass standard quantum limit at $\sim 40\,$Hz on its $40\,$kg silica test masses; depending on the precise modeling of the  parameter $a$ entering Eq.~\eqref{eq:DPscaling}, it is comparable to LISA Pathfinder's constraining power on $\sigma_{\rm DP}$ ~\cite{carlesso2016experimental}.

Additional bounds from spontaneous X-ray emission of electrons inside atoms place $\sigma_{\rm DP}\gtrsim 10\,\text{\AA}\sim 10^{-9}\,$m~\cite{aprile2025challenging}.  This numerically provides a bound superior to the SN scale, but the result here is obtained assuming that the DP force noise remains white for electrons up to frequencies beyond their Compton frequencies.  The Regime~I noise floor at the SN scale is of the same order as predicted signatures of the Causal-Conditional Schr\"odinger--Newton theory~\cite{liu2023semiclassical,liu2025semiclassical,liu2026testing,miki2025role} and the Correlated World-Line theory~\cite{wilson2022propagators,wilson2024testing}.  Experimental efforts toward those theories therefore double as tests of Regime~I.  Operationally, the requirement is to resolve an additional force noise at the level of the free-mass SQL at angular frequency $\omega_{\rm SN}$, which ranges from $2\pi \times 8\,{\rm mHz}$ for Si to $2\pi\times 78\,{\rm mHz}$ for Osmium~\cite{helou2017measurable}. 

As we concretely look at current experiments, for LISA Pathfinder, if we use $M=1.92\,$kg and $\omega_{\rm SN} = 0.28\,{\rm s}^{-1}$ (Platinum), the corresponding acceleration noise for Regime I is given by 
\begin{equation}
    \sqrt{S_a^{\rm I} }\approx 3\times 10^{-3}\,{\rm fm/s^2}/\sqrt{\rm Hz}
\end{equation}
while the best current acceleration noise, after accounting for known Brownian noise, is at $\sim 0.1\,{\rm fm/s^2}/\sqrt{\rm Hz}$~\cite{armano2024depth}. There is a gap of about 30 in amplitude. Space-based gravity-induced-entanglement (GIE) experiments, e.g., Ref.~\cite{matsumoto2025space}, may reach the requirement for probing Regime~I.

For a viscous mechanical loss with damping rate $\Gamma_m$ at temperature $T$, the thermal force noise is $S_F^{\rm th} = 4M\Gamma_m k_B T$; comparing with $S_F^{\rm CG} = 2\hbar M\omega_{\rm SN}^2$ gives the requirement
\begin{equation}
\Gamma_m \lesssim \frac{\hbar\omega_{\rm SN}^2}{2 k_B T} \approx 5\times 10^{-12}\,\mathrm{s}^{-1}\,\left(\frac{0.1\,\mathrm{K}}{T}\right)\left(\frac{\omega_{\rm SN}}{0.36\,\mathrm{s}^{-1}}\right)^2,
\label{eq:thermal}
\end{equation}
where $0.36\,\mathrm{s}^{-1}$ is the SN frequency of Tungsten~\cite{helou2017measurable}. This is far beyond the absolute damping rate of any current low-frequency mechanical sensor, but can still be a meaningful target for next-generation ultra-low-frequency torsion-balance and cryogenic suspension concepts~\cite{dai2024updating,ando2010torsion}, especially accounting for the development of new materials with lower dissipation~\cite{huang2025ultralow}.

\subsection{Local correlations  (Regime~II): noises and gravity induced entanglement}

In Sec.~\ref{sec:twoobjects} we showed that local auxiliary correlations within each ancilla does not prescribe a lower bound on single-object noise, and that the presence of a second object will then lead to noise that will prevent entanglement being generated in a way consistent with Refs.~\cite{kafri2014classical,fabiano2026minimal}.

In Regime~IIa, even though each object's noise floor $\hbar M\omega_{\rm CG}^2$ is not bounded from below,  as the second object is present, their mutual gravity force will add noise that cannot be made arbitrarily small by local squeezing.  For two comparable masses, the optimal local-sensing strength $\omega_{\rm CG}$ is at the scale of $\omega_G$, and hence the per-object noise floor for two comparable masses is shown be greater than
\begin{equation}
S_{f_G} \gtrsim 2\sqrt{2}\,\hbar M\omega_G^2,
\end{equation}
with $\omega_G \approx \sqrt{G\pi\rho/3}$ for two spherical bodies of density $\rho$ in contact, two orders of magnitude smaller than $\omega_{\rm SN}$ in amplitude.  This is at the scale of the free-mass Standard Quantum Limit for force measurement at $\omega_G$.  Since $\omega_G\ll\omega_{\rm SN}$ for ordinary solids, this force-noise target is numerically much lower than the Regime~I target; substituting into Eq.~\eqref{eq:thermal} the thermal-loss requirement tightens by $(\omega_{\rm SN}/\omega_G)^2 \sim 10^4$.  This is the level quoted by Refs.~\cite{kafri2014classical,fabiano2026minimal} for the LOCC bound, but here we would like to emphasize that it is not a fundamental bound on the noise of each object, but rather a consequence of the presence of the second object and the local-sensing structure. This provides a more nuanced understanding of the LOCC bound. Experimentally, by tuning the relative distance/orientations of objects, one can explore details of the additional classical gravity noise.

Regime~IIb is more general and is conceptually the most important part of the two-body analysis.  The noise of either individual object can be made small by local squeezing, but the joint force-noise matrix still obeys Eq.~\eqref{eq:det-bound-main}.  Thus correlated force noise is not enough to evade the no-entanglement result.  As long as the sensing remains local, the measurement-feedback channel is an LOCC channel and cannot generate gravitational entanglement.  Direct entanglement proposals such as Refs.~\cite{bose2017spin,marletto2017gravitationally} are therefore tests of whether nature is outside this local-sensing class; noise tomography of individual or single objects alone cannot falsify Regime~IIb, because the noise can be redistributed between the objects.

\section{Nonlocal correlations and linearized quantum gravity}
\label{sec:RIII}

We now turn to Regime~III, in which the auxiliary inputs associated with $A$ and $B$ are allowed to share nonlocal quantum correlations.  This is a qualitatively different resource from the local-record models of Secs.~\ref{sec:single} and~\ref{sec:twoobjects}: although the final feedback signals are still commuting records, the auxiliary observers is no longer a pair of independent local sensors.  Equivalently, one may formulate the same input--output structure using a common auxiliary system coupled to collective coordinates.  Such a nonlocal resource can generate entanglement between the masses.  We first demonstrate this in the instantaneous Newtonian limit and then ask when the construction can reproduce the response and fluctuations of a quantum gravitational field.

\subsection{Newtonian Gravity}

In the case of Newtonian gravity, in terms of the input fields, we can verify that the classical-gravity noise force operators commute:
\begin{equation}
    \label{eq:commute}
[\mathscr{F}_{A}, \mathscr{F}_{B}] = 0.
\end{equation}
Indeed, using $[a_{1\cdot G}(\omega), a_{2\cdot G}(\omega')] = i\delta(\omega-\omega')$ in Eqs.~\eqref{eq:cqa}--\eqref{eq:cqb}, the only potentially non-zero contributions are
\begin{equation}
\bigl[\alpha_{AG} a_{1AG},\, -\tfrac{\hbar K_{AB}}{\alpha_{AG}} a_{2AG}\bigr] + \bigl[-\tfrac{\hbar K_{AB}}{\alpha_{BG}} a_{2BG},\, \alpha_{BG} a_{1BG}\bigr] = 0,
\end{equation}
because the two terms are equal in magnitude and opposite in sign.
This commutativity is not specific to Regime~III; it also holds in the local-record regimes.  What distinguishes Regime~III is not the commutator itself, but the availability of cross-object quantum correlations in the auxiliary inputs.  An appropriately chosen joint input state can therefore squeeze the two commuting force noises simultaneously, reducing their joint noise below the local-channel determinant bound in Eq.~\eqref{eq:det-bound-main} and can generate entanglement between the masses.  In principle\,(though extremely difficult), the input ancilla states can be chosen so that the force noise can be arbitrarily close to zero, which represents a coherent, noiseless mutual Newtonian gravity interaction. Although the final feedback process still uses classical records, the auxiliary observer channel is no longer a local classical one: it functions operationally as a shared nonlocal resource or common mediator.  This is analogous to quantum teleportation, where pre-shared entanglement and classical communication together enable the transfer of quantum information~\cite{van2000multipartite}. In our model, the quantum resource in the ancilla is transferred to the two test masses. This is also related to the nonlocal entropic-force construction of Carney \emph{et al.}~\cite{carney2025quantum}. The retarded construction below shows how this nonlocal regime can be matched, at the level of linear response and correlation functions, to linearized quantum gravity; our result is also consistent with the general conclusion of Tomizuka and Takeda~\cite{tomizuka2026emergence}.

\subsection{Retarded response and the operational connection to linearized quantum gravity}

The possibility of realizing Newtonian gravity without additional noise also has to do with the fact that Newtonian gravity is instantaneous, and is described by a term in the Hamiltonian involving only the masses' position operators.  Analogously, the need for quantizing the electromagnetic field was not apparent in the Coulomb interaction between charges (which lead to the spectrum of the hydrogen atom), but only became apparent when non-classical photon statistics were confirmed~\cite{kimble1977photon}.

As Refs.~\cite{belenchia2018quantum,mitrakos2026does,christodoulou2023locally} have discussed, when the delay due to the finite speed of gravity becomes relevant, the quantum nature of the gravitational field becomes more apparent.  In particular, the existence of radiative gravitons and zero-point fluctuations of the gravitational field become intimately connected with the quantum nature of longitudinal Newtonian gravity.  This is because, when the delay is non-negligible, the commutator between forces on $A$ and $B$ does not vanish, and hence the gravity field must be able to accommodate information contents originally within masses $A$ and $B$ in order to mediate the interaction.  Furthermore, information flow from the gravity field to $A$ and $B$ will manifest as quantum fluctuations of gravity forces on $A$ and $B$. 

To be more concrete, let us consider positions of test masses coupled with a quantum field $\xi$:
\begin{equation}
    V =-\sum_J \hat x_J \hat \xi_J\,,
\end{equation}
where $\xi_J$ is the value of the field operator evaluated at the equilibrium position of mass $J$.  Using the language of linear response theory, we can write 
\begin{align}
    \hat x^{(1)}_J (t) &= \hat x^{(0)}_J (t) +\int^t_{-\infty} dt' \chi_{J}(t-t') \hat\xi_J^{(1)}(t') \\
    \hat \xi^{(1)}_J(t) &=  \hat \xi^{(0)}_J(t) +\int^t_{-\infty} dt' \sum_K G^{\rm ret}_{JK}(t-t')\hat x^{(1)}_K(t')
\end{align}
where $\chi_J$ is the response function of mass $J$ before coupling to the field and $G^{\rm ret}_{JK}$, the retarded Green Function, is the response function of the field $J$ to the position of the mass $K$, also given by the commutator
\begin{equation}
    G^{\rm ret}_{JK}(t-t') = i\left[\hat \xi_J^{(0)}(t),\hat \xi_K^{(0)}(t')  \right] \Theta(t-t') 
\end{equation}
where $\Theta(t-t')$ is the Heaviside step function.   We can then write 
\begin{equation}
    \left[\hat \xi_J^{(0)}(t),\hat \xi_K^{(0)}(t')  \right] = -i [ G^{\rm ret}_{JK}(t-t')- G^{\rm ret}_{KJ}(t'-t)] 
\end{equation}
Defining symmetrized and anti-symmetrized two-point functions
\begin{align}
    N_{JK}(t,t') =\frac{1}{2}
    \left\langle \left\{ \hat \xi_J(t),\hat \xi_K(t')\right\}\right\rangle \,,\\
        C_{JK}(t,t') =\frac{1}{2i}
    \left\langle \left[\hat \xi_J(t),\hat \xi_K(t')\right]\right\rangle \,,
\end{align}
Gaussian states are all those that satisfy the positive semi-definite condition: 
\begin{equation}
    N \pm  iC \succeq 0
\end{equation}
In the frequency domain, for symmetrized and single-sided spectral density, Gaussian states satisfy
\begin{equation}
    S_{IJ}^{\xi}(\Omega ) \pm 2 \mathrm{Im}G^{\rm ret}_{IJ}(\Omega) \succeq 0
\end{equation}

We note that the commutator vanishes when there is no time delay, which means no additional noise is necessary in order to realize Newtonian gravity. We also note that even for a single mass, this will lead to a minimum for the spectrum of the force noise: 
\begin{equation}
    S_\xi (\Omega)  \ge   2|\mathrm{Im} [G^{\rm ret} (\Omega)]|
    \end{equation}
This is the well-understood zero-point fluctuation associated with dissipation due to gravitational radiation, as previously discussed in the context of gravitational wave detectors\,\cite{Pang2018,Pang2019}.

\subsubsection{Realization by a feedback-control system}

We now argue that, even in the presence of propagation delay, Regime~III can reproduce the force noise required by a linear quantum field.  More broadly, a quantum-mediated interaction at the linear level can be represented by a linear monitoring-feedback system with an appropriate joint input state.

Let us assume a network of masses labeled by $I,J,\ldots$, with sensing strength $\alpha^J$, input quadratures $a_{1,2}^J$ and output quadratures $b_{1,2}^J$.  The force acting on mass $I$ is given by 
\begin{equation}
    \mathscr{F}_I (t) = \alpha_I a_{1}^I(t) +\int_{-\infty}^t dt' G^{\rm ret}_{IJ}(t-t') \frac{a_2^J(t')}{\alpha^J}
\end{equation}
where $G^{\rm ret}_{IJ}$ is the retarded response function of gravitational force on $I$ to displacement of object $J$.  Here the first term is back action from sensor $I$, while the second term is the sensing noise due to sensor $J$.  Evaluating the commutators between forces, we obtain 
\begin{equation}
\label{eqfcommgen}
[\mathscr{F}_{I}(t_1) ,\mathscr{F}_{J}(t_2)]=-iG_{IJ}^{\rm ret}(t_1-t_2) + iG_{JI}^{\rm ret}(t_2-t_1) \,.
\end{equation}
These are the same as the commutators of forces on the objects from the  quantum field whose Green function we are trying to achieve via monitoring and feedback.  Here we remind ourselves that we are using the Principle of Deferred Measurements and still treating these forces as Heisenberg Operators that depend on the output $b_2^I$ fields.  The equivalence here means we simply need to impose the same states for these forces in order to recover all the quantum-field predictions for linear systems.    More specifically, in steady Gaussian state, the symmetrized single-sided spectrum satisfies the same constraint as the quantum-field case:
\begin{equation}
    S_{IJ}^{\mathscr{F}}(\Omega) \pm 2 \mathrm{Im}G_{IJ}^{\rm ret}(\Omega) \succeq 0\,.
\end{equation}
As it turns out, the above equality can be achieved with input state of  $a_{1,2}^J$ that satisfy the Heisenberg Uncertainty relation. We can actually set $\alpha_I=1$, and write, in matrix form (suppressing $\Omega$ dependence):
\begin{equation}
    \mathbf{S}^{\mathscr{F}} = \mathbf{S}_{11} + \mathbf{S_{12}}\mathbf{G}^{\rm ret\,\dagger} +\mathbf{G}^{\rm ret} \mathbf{S}_{21}+\mathbf{G}^{\rm ret}\mathbf{S}_{22} \mathbf{G}^{\rm ret\,\dagger}\,.
\end{equation}
Decomposing $\mathbf{G}^{\rm ret}$ into real and imaginary parts,
\begin{equation}
    \mathbf{G}^{\rm ret} = \mathbf{A} + i\mathbf{K} \,,\quad \mathbf{A}=\mathbf{A}^\dagger,\;\quad \mathbf{K}=\mathbf{K}^\dagger\,, 
\end{equation}
in particular with $\mathbf{K}$ the imaginary part of $\mathbf{G}^{\rm ret}$  and writing
\begin{equation}
    \mathbf{S}_{22}=\mathbf{K}^{-1},\;
    \mathbf{S}_{12}=-\mathbf{A}\mathbf{K}^{-1},\; 
    \mathbf{S}_{21}=\mathbf{S}_{12}^\dagger,\;
    \mathbf{S}_{11} = \mathbf{A}\mathbf{K}^{-1} \mathbf{A}+\mathbf{K}, 
\end{equation}
We can show that 
\begin{equation}
    \begin{bmatrix}
        \mathbf{S}_{11} & \mathbf{S}_{12} \\
        \mathbf{S}_{21} & \mathbf{S}_{22}
    \end{bmatrix} 
    + i 
    \begin{bmatrix}
        0 & 1 \\
        -1 & 0
    \end{bmatrix} 
    \succeq 0
\end{equation}
and that 
\begin{equation}
\mathbf{S}^{\mathscr{F}}=2\mathbf{K}  =2\mathrm{Im}\mathbf{G}^{\rm ret}\,.
\end{equation}
For non-Gaussian states, the same commutation relation still allows the input field to be put onto any states allowed for the quantum field --- by matching all the higher-order correlation functions.

In other words, the input-output statistics of a linear quantum channel between position operators $x_J$ can be realized by linear monitoring and feedback supplied with correlated quantum inputs.  For non-Gaussian states, this equivalence additionally requires matching the relevant higher-order correlation functions.  The quantum information is then carried by the nonlocal resource state rather than created by classical communication; this is analogous to quantum teleportation, where pre-shared entanglement and classical communication together transfer a quantum state~\cite{van2000multipartite}.  Within its domain of validity, the construction can therefore represent non-Gaussian processes involving linearized gravitational modes, although implementing or observing such processes would be extraordinarily challenging.

\section{Conclusions}
\label{sec:conclusions}

The central goal of this paper has been to formulate a linear monitoring--feedback description of gravitational interactions that can describe semiclassical or classical-channel gravity and, in an extended regime, linearized quantum gravity within a unified framework.  We considered auxiliary observers/ancillas that monitor the positions of massive objects and feed their records back through a gravitational response kernel.  At the level of linear input--output theory, the deterministic response to external forces can be made identical to the target gravitational interaction.  The physically distinguishing information is carried by the auxiliary observers' noise, its cross-correlations, and its commutators.  This makes monitoring and feedback a useful language for comparing local-record models with nonlocal quantum-mediated models, and naturally provides a hierarchy for experiments targeted on testing the quantum nature of gravity.

The local-record monitoring--feedback class consists of models in which each object is monitored by its own auxiliary system, the feedback uses commuting classical records, and the input resources associated with distinct objects are mutually uncorrelated.  Regimes~I and II belong to this class: Regime~I has uncorrelated sensing and back action, while Regime~II allows correlations within each local ancilla.  Although the ancilla inputs may be nonclassical, the resulting gravitational interaction remains a local classical-record/LOCC channel.  Regime~III goes beyond this local-record class by allowing cross-object quantum correlations or a common nonlocal auxiliary system, providing an operational route to the linearized quantum-gravity limit.

Applying the framework to analyze the proposed experiment reveals a hierarchy. The most experimentally accessible consequence appears already for a single self-gravitating body.  In Regime~I, under uncorrelated/vacuum auxiliary monitoring, the additional self-gravity force noise satisfies
\begin{equation}
S_{\mathscr{F}}^{\rm min} = 2\hbar M\omega_{\rm SN}^2,
\end{equation}
with $\omega_{\rm SN}$ controlled by the microscopic mass distribution of the body.  The same scale appears in the mean-field Schr\"odinger--Newton equation studied by Yang \emph{et al.}~\cite{yang2013macroscopic}, but here as a noise floor rather than a self-attractive restoring frequency.  In a crystalline solid this scale is controlled by the localization of mass near lattice sites, rather than by the macroscopic size of the body, and the same construction connects parameter-freely to Di\'osi--Penrose collapse phenomenology through the scaling
\begin{equation}
\omega_{\rm DP} = {\omega_{\rm SN}}\,\Bigl(\frac{x_{\rm zp}}{\sigma_{\rm DP}}\Bigr)^{3/2}.
\end{equation}
The resulting force-noise target is numerically larger than the mutual-gravity scale in two-body entanglement proposals, and current mechanical bounds on the DP collapse model leave the SN scale within reach by roughly $(x_{\rm zp}/\sigma_{\rm DP})^{3/2}\sim$ a few in amplitude.  Single-body force-noise measurements are therefore a realistic first step in testing gravity models, even when a direct entanglement experiment is not yet sensitive enough.  This target does require treating the internal matter distribution as part of the classical record and specifying the local monitoring model; those assumptions are themselves part of what an experiment can constrain.

For two mutually gravitating masses, the analysis clarifies the role of correlations.  In Regime~I the auxiliary inputs are uncorrelated and each object retains essentially the single-body SN-scale noise floor, with only $O(\omega_G^2/\omega_{\rm SN}^2)\sim 10^{-4}$ corrections.  In Regime~II, local correlations can redistribute the noise and even suppress the apparent noise on one object; nevertheless, as long as the ancillas remain independent across objects, the joint force-noise matrix satisfies the determinant bound
\begin{equation}
S_{AA}\,S_{BB}-|S_{AB}|^2 \;\ge\;4 \hbar^2 K_{AB}^2,
\end{equation}
which is the local classical-channel/no-entanglement bound of Kafri--Taylor~\cite{kafri2014classical} and Fabiano \emph{et al.}~\cite{fabiano2026minimal}.  Thus, correlated force noise alone is not evidence for a quantum mediator: within the local-record class it can arise from ordinary feedback geometry and local quantum correlations, while entanglement remains forbidden.  Regime~IIa gives a concrete distance- and orientation-dependent noise target at the scale of $\sim\hbar M\omega_G^2$, whereas Regime~IIb shows that the noise may be redistributed between the two objects without removing the joint determinant constraint.

The precise bound relevant to an experiment depends on the composition of the masses, the mass of the companion object, and their separation.  If one additionally requires an {\it intrinsic} monitoring rate tied to each microscopic constituent, rather than allowing the monitoring strength to be optimized independently for each configuration, not all noise inequalities need be saturated; this restriction could provide a stronger bound on local-record models.

Regime~III is the essential extension beyond local-record models.  If the auxiliary inputs associated with the two masses share nonlocal quantum correlations, or if the auxiliary sector is common to the two masses, cross-object quantum resources can squeeze the commuting force combinations and generate gravity-induced entanglement.  More generally, for a prescribed linear retarded response $G^{\rm ret}$, the correlated input state can be chosen so that the monitoring--feedback forces have the same commutators and symmetrized spectra as the force operators of the corresponding quantum field.  In this operational linear sense, linearized quantum gravity can be reproduced by monitoring and feedback, including non-Markovian fluctuations and, in principle, non-Gaussian input states.  The statement is about equivalence of the linear response and correlation functions; it does not assert equivalence to the full nonlinear or ultraviolet theory of quantum gravity.

Our framework provides a unified description of linear gravitational interactions in terms of their response functions and force-noise correlations, connecting several approaches to gravitational dynamics. In the instantaneous Newtonian limit, entropic-force models can reproduce the coherent Newtonian interaction while introducing additional diffusion through their auxiliary mediators~\cite{carney2025quantum,fabiano2026minimal}. These models can be compared with the force-noise predictions of Regimes I and II, where independent local monitoring introduces additional noise and the local classical channel's bound forbids gravity-induced entanglement. In contrast, nonlocal mediator constructions are more closely related to Regime III, in which shared quantum resources can allow entanglement.

The retarded response and force-noise correlations in Regime III can also cover the gravitational dissipation arising from different physical mechanisms. In linearized gravity, gravitational-wave emission and the associated radiation reaction give rise to dissipation, as investigated in Refs.~\cite{Pang2018,Pang2019}. Alternatively, gravitational dissipation may originate from a temporally nonlocal modification of the gravitational response, as proposed in Ref.~\cite{yang2015}. Such a nonlocal response may also emerge as an effective four-dimensional description after integrating out additional degrees of freedom, including those propagating along an extra dimension. Furthermore, the state dependence of gravitational fluctuations has been investigated in Refs.~\cite{parikh2020,parikh2021}. Within the linearized detector dynamics, these fluctuations can be characterized by the corresponding noise correlations. 

These connections highlight two distinct benchmarks for gravitational interactions: the no-entanglement constraint imposed by LOCC and the vacuum-fluctuation constraint associated with quantum fields. Measurements of coherent response, force-noise correlations, and entanglement therefore provide complementary means of probing gravitational dynamics and its possible quantum nature.

\begin{acknowledgments}

We acknowledge useful discussions with Dan Carney and Hiroki Takeda. 
Research of D.M.\ is supported by the JSPS and the Keck Foundation. Research of A.L.\ and Y.C.\ is supported by the Simons Foundation (Award Number 568762).  Research of Y.M.\ is supported by the Natural Science Foundation of China under Grants No. 12474481, No. 12441503.

\end{acknowledgments}

\appendix
\section{No-entanglement determinant bound in Regime~IIb}
\label{app:noent}

We make the no-entanglement statement of Regime~IIb explicit by deriving the determinant bound \eqref{eq:det-bound-main}.  Let $K\equiv K_{AB}$, $\alpha_A\equiv \alpha_{AG}$, $\alpha_B\equiv\alpha_{BG}$.  In Regime~IIb the renormalized force-noise operators are
\begin{align}
f_A &= \alpha_A\,v_{1A} - \frac{\hbar K}{\alpha_B}\,v_{2B}, \quad
f_B = \alpha_B\,v_{1B} - \frac{\hbar K}{\alpha_A}\,v_{2A},
\end{align}
with $v_{1A,2A}, v_{1B,2B}$ dimensionless quadratures normalized as in \eq{eq:spectral-convention}.  Let $A_1 \equiv S^A_{v_1v_1}$, $A_2 \equiv S^A_{v_2v_2}$, $A_{12}\equiv S^A_{v_1v_2}$, and similarly $B_1, B_2, B_{12}$ for the $B$-ancilla.  Since there are no cross-$A$-$B$ correlations,
\begin{align}
S_A &= \alpha_A^2 A_1 + ({\hbar^2 K^2}/{\alpha_B^2})B_2\,,\\  S_B &= \alpha_B^2 B_1 + (\hbar^2 K^2/\alpha_A^2)\,A_2,\\
S_{AB} &= -\hbar K\bigl(A_{12} + B_{12}\bigr).
\end{align}
Local Heisenberg uncertainty in the convention of \eq{eq:spectral-convention} gives $A_1 A_2 - |A_{12}|^2 \ge 1$ and $B_1 B_2 - |B_{12}|^2 \ge 1$.  For real symmetrized cross spectra, direct expansion yields
\begin{align}
S_A S_B - S_{AB}^2 ={}& \alpha_A^2\alpha_B^2 A_1 B_1 + \frac{\hbar^4 K^4}{\alpha_A^2\alpha_B^2} A_2 B_2 \nonumber\\
&{} + \hbar^2 K^2\bigl(A_1 A_2 + B_1 B_2\bigr) - \hbar^2 K^2 (A_{12}+B_{12})^2.
\end{align}
The arithmetic--geometric-mean inequality applied to the first two terms gives
\(\alpha_A^2\alpha_B^2 A_1B_1 + (\hbar^4 K^4/\alpha_A^2\alpha_B^2) A_2B_2 \ge 2\hbar^2 K^2\sqrt{A_1A_2 B_1 B_2}\);
combined with $A_1A_2 + B_1B_2 \ge 2\sqrt{A_1 A_2 B_1 B_2}$, this gives
\begin{equation}
S_A S_B - S_{AB}^2 \;\ge\; \hbar^2 K^2\!\left[\bigl(\sqrt{A_1A_2}+\sqrt{B_1B_2}\bigr)^2 - (A_{12}+B_{12})^2\right].
\end{equation}
Using $\sqrt{A_1A_2}\ge\sqrt{A_{12}^2 + 1}$ and similarly for $B$, together with $\sqrt{(A_{12}^2+1)(B_{12}^2+1)} \ge |A_{12}B_{12}| + 1$, one obtains $(\sqrt{A_1A_2}+\sqrt{B_1B_2})^2 - (A_{12}+B_{12})^2 \ge 4$; hence
\begin{equation}
S_A S_B - S_{AB}^2 \;\ge\; 4\hbar^2 K^2.
\end{equation}

\bibliographystyle{apsrev4-2}
\bibliography{main}

\end{document}